\documentclass[11pt,aip,preprint,tightenlines]{revtex4-2}
\usepackage{amsmath}
\usepackage{amssymb}
\usepackage{bm}
\usepackage[mathscr]{eucal}
\usepackage{graphicx}
\usepackage{psfrag}
\usepackage{subfigure}
\usepackage{color}
\usepackage{wasysym}
\include{graphix}
\usepackage{framed}
\usepackage{enumitem}%
\usepackage{lipsum}
\usepackage{float}
\usepackage{stmaryrd}
\usepackage{MnSymbol}
\definecolor{orange}{rgb}{1,0.5,0}
\usepackage[normalem]{ulem}
\usepackage{cancel}

\usepackage{tikz}

\usetikzlibrary{patterns}
\usetikzlibrary{shapes}

\graphicspath{{figs/}}

\newcommand{\myqed}{\nobreak \ifvmode \relax \else
      \ifdim\lastskip<1.5em \hskip-\lastskip
      \hskip1.5em plus0em minus0.5em \fi \nobreak
      \vrule height0.75em width0.5em depth0.25em\fi}

\newcommand{\mathd}{\mathrm{d}}

\newcommand{\myRe}{\mathrm{Re}}
\newcommand{\myWe}{\mathrm{We}}
\newcommand{\myBo}{\mathrm{Bo}}
\newcommand{\myCn}{\mathrm{Cn}}
\newcommand{\myPe}{\mathrm{Pe}}
\newcommand{\mydiff}{\mathcal{D}}

\newcommand{\lam}{R}

\newcommand{\chempot}{\Phi}
\newcommand{\surftenx}{\widehat{\sigma}}
\newcommand{\surften}{\sigma}

\newcommand{\vecu}{\bm{u}}
\newcommand{\vecx}{\bm{x}}

\newcommand{\hatnwall}{\widehat{\bm{n}}_f}
\newcommand{\thetaeq}{\vartheta_0}

\newcommand{\betamax}{\beta_{max}}

\newcommand{\hatz}{\widehat{\bm{e}}_3}

\newcommand{\dittotikz}{%
    \tikz{
        \draw [line width=0.12ex] (-0.2ex,0) -- +(0,0.8ex)
            (0.2ex,0) -- +(0,0.8ex);
        \draw [line width=0.08ex] (-0.6ex,0.4ex) -- +(-1.5em,0)
            (0.6ex,0.4ex) -- +(1.5em,0);
    }%
}

\begin{document}

\title{Analysis of the Spreading Radius in Droplet Impact: The Two-Dimensional Case}
\author{Lennon \'O N\'araigh}
\email{Corresponding author.  Email: onaraigh@maths.ucd.ie}
\affiliation{School of Mathematics and Statistics, University College Dublin, Belfield, Dublin 4, Ireland}
\author{Juan Mairal}
\affiliation{ I3A,  University of Zaragoza, Zaragoza, Spain}

\date{\today}

\begin{abstract}
We study droplet-impact problems in a three-dimensional cylindrical or equivalent two-dimensional Cartesian geometry.  Such structures do have an approximate experimental realization, and they are often simulated  a test-bed for computational methods.  We focus on droplet impact on a smooth homogeneous surface as well as head-on collision of two droplets.
We perform an energy-budget analysis and introduce a correlation which predicts the maximum spreading radius as a function of Reynolds number and Weber number.  We show how the dissipation term in this analysis can be decomposed into boundary-layer dissipation in the droplet lamella (where applicable), and head loss.   We use existing results  in the literature (simulations and experiments), as well as our own simulation results to validate the correlation.  Dissipation by head loss is a key term in the analysis: only by modeling it accurately can one obtain good agreement between the simulations and the theory.
\end{abstract}

\maketitle

\section{Introduction}
\label{sec:intro}

The impact of a droplet of fluid on a smooth, homogeneous substrate is a well studied problem, with many practical applications, e.g. inkjet printing~\cite{yarin2006drop}, cooling~\cite{yarin2006drop,valluri2015}, and crop spraying~\cite{yarin2006drop,Moghtadernejad2020}. 
Different impact regimes occur depending on the droplet's Weber number and Reynolds number.  In particular, there is a splashing threshold $\myWe\sqrt{\myRe}\apprle 3,000$, below which the impacting droplet retains an axisymmetric shape and spreads across the substrate until a maximum spreading radius is attained, whereupon retraction occurs~\cite{josserand2016drop}.  Naturally, this is a three-dimensional (albeit axisymmetric) phenomenon.    However, in a recent experiment by Le Jeune et al.~\cite{lejeune2018edge}, rectangular liquid sheets have been created after an asymmetric droplet impact.  In a more recent experiment  by N\'eel et al.~\cite{neel2020fines}, the authors created elongated liquid cylinders, which were then subjected to a head-on collision.   In this second experiment, the resulting spreading phenomenon is qualitatively different to what would be observed in a 3D axisymmetric head-on collision.  As such, a theoretical analysis of droplet impact and droplet spread in the case of liquid cylinders may be useful to understand these experiments.
Below a critical Reynolds number and Weber number (analogous to the splash threshold in three dimensions~\cite{neel2020fines}),  such impact phenomena can be treated as quasi-two-dimensional.  Hence, the proposed theoretical model in the present work considers a two-dimensional droplet.  This is an idealization but it can be thought of as approximating a cylindrical droplet impact below the splash threshold.


 A second motivation is that many computational studies have been performed in the case of droplet impact and droplet spread for such (idealized) 
two-dimensional systems~\cite{ding2007diffuse,shin2009simulation,gupta2011two,wu2017dynamics,wu2021decoupled,rafi2022two}.   These simulations have been performed as test-beds either for the analogous but more complex 3D axisymmetric geometry, or for fully 3D non-axi-symmetric geometry.  As part of these test simulations, the authors analyze the extent to which the droplet spreads after impact, as a function of time.  Maximum spreading occurs just before retraction.  The maximum spreading radius is a function of Weber number and Reynolds number.  
Analogous to the three-dimensional case, Gupta and Kumar~\cite{gupta2011two} develop a theoretical correlation for the maximum spreading radius as a function of Reynolds number, Weber number, and contact angle.  The correlation is based on an energy-budget analysis.

We are motivated to further consider this theoretical approach by an analogy with the 3D  axisymmetric case.  There, an energy-budget analysis for the maximum spreading radius is known to over-predict the maximum spreading radius in the rim-lamella regime~\cite{wildeman2016spreading}.  This refers to a regime at large Weber number and intermediate Reynolds number where in the extremity of the spreading droplet forms a `bulge' or rim, the dynamics of which are governed by surface tension.  In this case, a rim-lamella model~\cite{eggers2010drop} (often coupled with an energy-budget analysis~\cite{roisman2002normal}) is used to predict the maximum spreading radius.  Motivated by these observations,  the main aim of the present work is to develop an energy-budget analysis in the case of 2D droplets (or equivalently, 3D cylindrical droplets), and to correct for the rim.  In this way, we will develop predictions for the maximum spreading radius and compare these to simulation data and  experiments.

This paper is organized as follows.  In Section~\ref{sec:theory} we introduce a theoretical model for the maximum spreading radius in 2D droplet impact problems.
As we are interested in comparing the theoretical model with evidence from simulations, we present a method for generating such simulations in Section~\ref{sec:numerical}, based on the Diffuse Interface Method.  We compare the theoretical model with our own simulation results in case of droplet impact on a smooth surface in Section~\ref{sec:impact}.  We  compare the predictions of the theoretical model with simulation results (both our own and other results drawn from the literature) in case of head-on droplet collision in Section~\ref{sec:head-on}.  We also compare the predictions of the theoretical model with simulation results.  
 Discussion and concluding remarks are presented in Section~\ref{sec:conc}.
Throughout the work, we use the following definitions for Weber number and Reynolds number:
	\begin{equation}
	\myWe=\frac{\rho U_0^2 R_0}{\sigma},\qquad \mathrm{Re}=\frac{\rho U_0 R_0}{\mu},
	\label{eq:Wedef}
	\end{equation}
	where $\rho$ is the fluid density, $\sigma$ the surface tension,   and $\mu$ is the fluid viscosity.  Also, $U_0$ is the droplet's speed prior to impact, and $R_0$ is the radius of the  droplet prior to impact.  We use the notation $r_{max}$ to denote the maximum spreading radius of the droplet after impact, and the variable $\betamax=r_{max}/R_0$ to denote the non-dimensional maximum spreading radius.

\section{Theoretical Model}
\label{sec:theory}

In general, there is no analytical formula for $r_{max}(\myWe,\myRe)$.  In the case of 3D axisymmetric droplets, a number of correlations exist in the literature.  At least three approaches to formulating the correlation exist.  The first  involves semi-empirical formulae based on a very wide range of experimental results, such as the work by Roisman~\cite{roisman2009inertiaP2}, where the 3D leading-order scaling $r_{max}\sim \myRe^{1/5}$ is elucidated, in case of the viscous-spreading regime ($P>1$, $P=\myWe/\myRe^{4/5}$).  A second approach involves an energy-budget analysis, wherein the pre-impact energy of the droplet is equated to the energy at maximum spreading, plus the energy dissipated~\cite{chandra1991collision,wildeman2016spreading}.   A third approach, based on momentum balances, is valid in the rim-lamella regime of droplet spreading~\cite{roisman2002normal,eggers2010drop}.  All approaches generate equivalent scaling behaviors in the appropriate asymptotic limits.  For the present purposes, we use the energy-budget analysis, tailored to 2D droplets, and pay particular attention to the modeling of the energy dissipation, which is key to producing a correlation that matches the available experimental and simulation data.


\subsection{Energy-Budget Analysis}

For a cylindrical drop with pre-impact speed $U_0$ and radius $R_0$, the initial energy of the system is:
\begin{equation}
E_{init}=\left[\tfrac{1}{2}\rho (\pi R_0^2)U_0^2+2\pi R_0\surften\right] \lambda,
\end{equation}
where $\lambda$ is the length of the cylindrical droplet in the third dimension, this is an irrelevant variable as it cancels from both sides of the following energy-budget analysis.   At maximum spreading, it is assumed that the droplet's internal kinetic energy is negligible, such that the droplet's energy can be attributed entirely to surface energy:
\begin{equation}
E_{final}=\surften \left(\Pi-2r_{max}\cos\vartheta\right) \lambda.
\end{equation}
Here, $\Pi$ is the perimeter of the droplet at maximum spread, and $\vartheta$ is the advancing contact angle.
The initial and final energies are related through an energy balance:
\begin{equation}
E_{init}=E_{final}+\Delta E,
\label{eq:eb1}
\end{equation}
where $\Delta E$ denotes the energy loss through viscous dissipation.  

According to Wildeman et al.~\cite{wildeman2016spreading}, the energy loss due to viscous dissipation is made up of at least two parts.  The first part is viscous dissipation due to a boundary layer which forms inside the droplet lamella. The thickness of the boundary layer is estimated as $\delta \sim \sqrt{\nu \tau}$, where $\nu$ is the liquid kinematic viscosity and $\tau$ is the time interval between the formation of the boundary layer and the time of maximum droplet spread.  Hence, $\Delta E_{bl}\sim \mu(U_0/\delta)^2[\delta (2r_{max})\lambda]\tau$. We estimate $\tau \approx (r_{max}/U_0)-(R_0/U_0)$, that is, the time for maximum spread to occur, minus a time interval $R_0/U_0$ of an initial phase prior to the onset of the boundary layer.   Hence, the energy loss due to the boundary layer is estimated as:
\[
\Delta E_{bl}\approx  \frac{2a}{\sqrt{\myRe}}\rho_L U_0^2 \beta_{max}\sqrt{\beta_{max}-1}(R_0^2\lambda),
\]
where $a$ is an $O(1)$ constant, and $\beta_{max}=r_{max}/R_0$.

A second source of dissipation is so-called `head loss', this occurs during a sudden expansion of a flow passing from a narrow channel into a wider channel, this is analogous to the scenario that occurs when the flow in the lamella passes into the rim.  According to Wildemen et al.~\cite{wildeman2016spreading} and Villermaux and Bossa~\cite{villermaux2011drop}, the head loss can be estimated as a simple fraction of the initial kinetic energy of the droplet prior to impact.  As viscous dissipation prior to impact is negligible, the kinetic energy of the droplet prior to impact corresponds almost exactly to the initial potential energy of the droplet, $(1/2)\rho(\pi R_0^2)U_0^2\lambda$.  Thus, the head loss is estimated as 
\begin{equation}
\Delta E_{h}\approx  b \left[\tfrac{1}{2}\rho (\pi R_0^2)U_0^2\lambda\right],
\label{eq:head}
\end{equation}
where $0<b<1$ is another constant.
In this way, the energy budget~\eqref{eq:eb1} becomes:
\begin{multline}
\tfrac{1}{2}\rho (\pi R_0^2)U_0^2+2\pi R_0\surften =
\surften \left(\Pi-2r_{max}\cos\vartheta\right) \\
+\frac{2a}{\sqrt{\myRe}}\rho_L U_0^2 \beta_{max}\sqrt{\beta_{max}-1}\,R_0^2
+b \left[\tfrac{1}{2}\rho (\pi R_0^2)U_0^2\right].
\label{eq:energyb}
\end{multline}

A third loss channel active at low Weber number has been identified by Park et al.~\cite{park2003single} and Huang and Chen~\cite{huang2018energetic}.  This is the phenomenon of spontaneous surface dissipation.  This is absent from previous studies at high Weber number (e.g. by Wildeman et al.~\cite{wildeman2016spreading}, Roisman et al.~\cite{roisman2002normal}), but is relevant in the present study. However, this loss channel can be incorporated into the expression for head loss: as such, Equation~\eqref{eq:head} represents energy losses through all channels other than boundary-layer dissipation.

We now divide Equation~\eqref{eq:energyb} across by $(1/2)\rho (\pi R_0^2)U_0^2$ to obtain:
%
\begin{equation}
(1-b)+\frac{4}{\myWe}  =
\frac{2}{\pi\myWe}\left(\frac{\Pi}{R_0}-2\beta_{max}\cos\vartheta\right) \\
+\frac{4}{\pi}\frac{a}{\sqrt{\myRe}}\beta_{max}\sqrt{\beta_{max}-1}.
\label{eq:eb2}
\end{equation}
%
To make further progress, it is necessary to have an expression for the droplet perimeter $\Pi$.  A rough approximation can be made by treating the droplet as a rectangle.  The equivalent three-dimensional approximation would be a disk, this is a standard approach in the literature~\cite{chandra1991collision}.  The droplet area is fixed as $A_0=\pi R_0^2$, this is then equal to the droplet area at maximum spread, $A_0=2r_{max}h$, where $h$ is the height of the droplet at maximum spread.  Hence, the perimeter is approximated as $\Pi=2r_{max}+A_0/r_{max}$.      Thus, Equation~\eqref{eq:eb2} becomes:
\begin{equation}
(1-b)+\frac{4}{\myWe}   =
\frac{2}{\pi\myWe}\left[2\beta_{max}\left(1-\cos\vartheta\right)+ \frac{\pi}{\beta_{max}}\right]
+\frac{a}{\sqrt{\myRe}}\beta_{max}\sqrt{\beta_{max}-1}.
\label{eq:correlation}
\end{equation}
(the factor of $4/\pi$ in the dissipation term has been absorbed into $a$).
Equation~\eqref{eq:correlation} does not have an explicit solution for $\beta_{max}$ in terms of $\myWe$ and $\myRe$, however, it can be solved numerically for the purpose of comparing with experiments and simulations.  Furthermore, explicit solutions exist in certain limiting cases.

\subsection{Asymptotic Limits}

We now examine some special limiting cases of Equation~\eqref{eq:correlation} where an explicit expression for $\betamax$ is possible.

\paragraph*{Inviscid Limit:} For  $\myRe\rightarrow\infty$, Equation~\eqref{eq:correlation} reduces to
\begin{equation}
\frac{\pi}{2}(1-b)\myWe+2\pi=\left[2\beta_{max}(1-\cos\vartheta)+\frac{\pi}{\beta_{max}}\right],
\label{eq:inviscid}
\end{equation}
with exact solution
\[
\betamax=\frac{\omega+\sqrt{\omega^2-8\pi(1-\cos\vartheta)}}{\tfrac{2}{\pi}(1-\cos\vartheta)},\qquad \omega=\frac{\pi}{2}(1-b)\myWe+2\pi.
\]
For $\myWe$ large but finite, this further reduces to:
\[
\beta_{max}\approx \frac{\myWe\,\pi(1-b)}{4(1-\cos\vartheta)}.
\]
The equivalent scaling behavior for 3D axisymmetric droplets is:
\[
\beta_{max}\approx \sqrt{\frac{4}{1-\cos\vartheta}\left[\tfrac{1}{12}(1-b)\myWe+1\right]},
\]
hence, $\beta_{max}\sim \myWe$ for cylindrical droplets and $\beta_{max}\sim \myWe^{1/2}$ for axisymmetric droplets.

\paragraph*{Finite viscosity, large Weber number:} For $\myWe\rightarrow\infty$ Equation~\eqref{eq:correlation} reduces to:
\[
\frac{\pi}{2}(1-b)\approx \frac{2a}{\sqrt{\myRe}}\beta_{max}\sqrt{\beta_{max}-1},
\]
For $\myRe$ large but finite, this gives $\beta_{max}\sim \myRe^{1/3}$.  The corresponding result for 3D axisymmetric droplets is $\beta_{max}\sim \myRe^{1/5}$.


\subsection{Head-On Collision of two droplets}

The head-on collision of two 2D droplets has been investigated by Wu et al.~\cite{wu2021decoupled}  In a previous study~\cite{willis2003binary}, in case of 3D axisymmetric droplets, it is argued that the head-on collision of two such droplets is equivalent (in terms of energy-budget analysis) to to a droplet impacting on a surface with a free-slip boundary condition and a contact angle of $\pi/2$.  In this case, there is no viscous boundary layer.  As such, an appropriate correlation for the head-on collision of two droplets in 2D is simply Equation~\eqref{eq:inviscid}, with $\vartheta=\pi/2$:
\begin{equation}
\frac{\pi}{2}(1-b)\myWe+2\pi=\left(2\beta_{max}+\frac{\pi}{\beta_{max}}\right).
\label{eq:inviscid1}
\end{equation}

\section{Numerical Method}
\label{sec:numerical}

The aim of the present work is to validate the correlations developed in Section~\ref{sec:theory}, using data from simulations and experiments.  We use simulation data from the literature, supplemented with our own.  For the latter purpose, we use a diffuse interface method to model droplet impact and droplet collisions.  We summarize the methodology below.

\subsection{Diffuse-Interface Method}

We use the Diffuse-Interface Method to describe the interface separating the gas from the liquid.  A phase-field function  $C(\vecx,t)$ is introduced, with (conventionally) $C=1$ indicating the liquid phase and $C=-1$ the surrounding gas phase.  Thus, $C(\vecx,t)=1$ if the region around the point $\vecx$ is rich in liquid phase at time $t$, and $C(\vecx,t)=-1$ if the region around the same point is rich in the gas phase.  Also, the level set $C=0$ represents a demarcation between the liquid and the gas phases.  In this approach, the energy  associated with interfaces is estimated from a diffuse-interface functional:
\begin{equation}
F[C]=\frac{\surftenx}{\xi}\int_\Omega \left[f_0(C)+\tfrac{1}{2}\xi^2|\nabla C|^2\right]\mathd^3 x.
\label{eq:free}
\end{equation}
Here, $\surftenx$ is a constant with units of surface tension, and $\Omega$ is the fluid domain.  
The evolution for $C$ is given by:
\begin{subequations}
\begin{equation}
\frac{\partial C}{\partial t}+\vecu\cdot\nabla C=\nabla\cdot\left[M(C)\nabla\chempot\right],
\label{eq:dimC}
\end{equation}
where $M(C)\geq 0$ is a non-negative mobility function, and 
\begin{equation}
\chempot=\delta F/\delta C=\frac{\surftenx}{\xi}\left[f_0'(C)-\xi^2\nabla^2 C\right].
\end{equation}%
\label{eq:modelCH}%
\end{subequations}%

The Diffuse-Interface Method is essentially a `one-fluid' formulation.  As such, there is a single fluid velocity $\vecu(\vecx,t)$, a single fluid density, and a single fluid viscosity.  However, the fluid density $\rho$ and fluid viscosity $\mu$ depend on the  amount of each fluid phase present at a particular point:
\begin{subequations}
\begin{eqnarray}
\rho&=&\rho_G \left[\tfrac{1}{2}(1-C)\right]+\rho_L\left[\tfrac{1}{2}(1+C)\right],\\
\mu&=&\mu_G \left[\tfrac{1}{2}(1-C)\right]+\mu_L\left[\tfrac{1}{2}(1+C)\right].
\end{eqnarray}%
\label{eq:rhomu}%
\end{subequations}
Here, $\rho_L$ and $\rho_G$ are the constant densities in each of the pure fluid phases (gas and liquid in the present context) and $\mu_L$ and $\mu_G$ are the constant dynamic viscosities in the same.

The velocity field $\vecu(\vecx,t)$ then satisfies a Navier--Stokes equation with a Korteweg stress term:
\begin{equation}
\rho\left(\frac{\partial\vecu}{\partial t}+\vecu\cdot\nabla\vecu\right)=-\nabla p+\nabla\cdot\left[\mu\left(\nabla\vecu+\nabla\vecu^T\right)\right]+\chempot\nabla C+\rho\bm{g}.
\label{eq:dimNS}
\end{equation}
The Korteweg stress term is $\chempot\nabla C$, this describes the effect of surface tension in the diffuse-interface formalism.  Here also, $\bm{g}$ is the acceleration due to gravity.

Lastly, we comment on the choice of interpolation in Equation~\eqref{eq:rhomu}, as this is influenced by the mobility function $M(C)$.  This is often taken to be $C$-dependent, this guarantees $|C|\leq 1$.  Following Ewetola  et al.~\cite{ewetola2021control}, and for simplicity, we take the mobility $M(C)$ to be a constant.  This choice occasionally produces values of $C$ such that $|C|>1$.  When this occurs, the expressions~\eqref{eq:rhomu} are modified, e.g. the expression for the density becomes
\[
\rho=\rho_G \min\{-1,\left[\tfrac{1}{2}(1-C)\right]\}+\rho_L\max\{1,\left[\tfrac{1}{2}(1+C)\right]\};
\]
a similar expression then applies to the viscosity.

Summarizing, the Diffuse-Interface model consists of four equations: Equation~\eqref{eq:modelCH} for the phase field $C$,  Equation~\eqref{eq:rhomu} for interpolating the density and viscosity between the different fluid phases, Equation~\eqref{eq:dimNS} for the velocity field $\vecu$, and finally, a mass-conservation equation, which we take to be the incompressibility condition
\begin{equation}
\nabla\cdot\vecu=0.
\label{eq:dimDiv}
\end{equation}
The equations of motion for the Diffuse Interface Method are implemented using an established in-house  computational framework based on prior works~\cite{naraigh2014linear,naraigh2015flow,fannon2016high}.

Mass conservation $(d/dt)\int_{\Omega}C\mathd^3 x=0$ is further ensured by taking $\hatnwall\cdot\nabla \Phi=0$ on $\partial\Omega$, the boundary of the domain $\Omega$, here $\hatnwall$ is the outward-pointing unit normal to $\partial\Omega$.  Also, the no-slip condition $\bm{u}=0$ is applied on $\partial\Omega$.  However, because Equation~\eqref{eq:modelCH} is a fourth-order partial-differential equation, a further boundary condition is needed.  The  freedom afforded by the extra boundary condition enables us to prescribe the behaviour of the contact line.  On boundary faces where a contact line exists, we impose the so-called \textit{geometric boundary condition}~\cite{ding2007wetting}:
\begin{equation}
\hatnwall\cdot\nabla C=-\tan\left(\tfrac{1}{2}\pi-\vartheta\right)\left|\nabla C-\left(\hatnwall\cdot\nabla C\right)\hatnwall\right|,
\label{eq:dimGeom}
\end{equation}
where $\thetaeq$ is the static contact angle.  Using this geometric condition gives the same results as other boundary conditions based on free-energy arguments~\cite{ding2007wetting,ewetola2021control}.

\subsection{Choice of Diffuse-Interface parameters} 

We use the Diffuse-Interface Method to model droplet impact and droplet spread on a smooth surface.  In dimensionless variables, the Cahn--Hilliard equation~\eqref{eq:modelCH} describing the interface location becomes:
\begin{equation}
\frac{\partial C}{\partial t}+\vecu\cdot\nabla C=\mydiff\nabla^2\left[f_0'(C)-\myCn^2\nabla^2C\right],\qquad \mydiff=\frac{R_0}{\xi} \frac{M_0\surftenx}{UL^2},\qquad \myCn=\xi/R_0,
\label{eq:modelCH1}
\end{equation}
where $M_0\geq 0$ is the constant mobility.  We identify the Cahn number $\myCn=\xi/R_0$ and the interfacial P\'eclet number $\myPe=UL^2/(M_0\surftenx)$. 
The coefficient $\surftenx$ is related to the surface tension.  This statement can be made precise through the equation
\begin{equation}
\surften=\xi\surftenx\int_{-\infty}^\infty \left(\frac{\mathd C_*}{\mathd z}\right)^2\mathd z,
\label{eq:theory}
\end{equation}
where we have reverted momentarily to the dimensional variables.  Here $C_*=\tanh(z/\sqrt{2}\xi)$ is the equilibrium one-interface solution to the Cahn--Hilliard equation in the absence of flow.  The justification for Equation~\eqref{eq:theory} was given by Ding and Spelt~\cite{ding2007diffuse}.  Using Equation~\eqref{eq:theory}, we obtain $\surftenx=(3/2\sqrt{2})\surften$.  Thus, the dimensionless Navier--Stokes equations with surface tension now read:
\begin{subequations}
\begin{equation}
\widehat{\rho}\left(\frac{\partial\vecu}{\partial t}+\vecu\cdot\nabla\vecu\right)=-\nabla p+\frac{1}{\myRe}\nabla\cdot\left[\widehat{\mu}\left(\nabla\vecu+\nabla\vecu^T\right)\right]\\+
\tfrac{3}{2\sqrt{2}}\frac{1}{\myCn}\frac{1}{\myBo}\left[f_0'(C)-\myCn^2\nabla^2 C\right]\nabla C-\widehat{\rho}\,\hatz
\label{eq:dimNS_new}
\end{equation}
where
\begin{equation}
\widehat{\rho}=\frac{\rho_G}{\rho_L} \min\{-1,\left[\tfrac{1}{2}(1-C)\right]\}+\max\{1,\left[\tfrac{1}{2}(1+C)\right]\},
\end{equation}
and
\begin{equation}
\widehat{\mu}=\frac{\mu_G}{\mu_L} \min\{-1,\left[\tfrac{1}{2}(1-C)\right]\}+\max\{1,\left[\tfrac{1}{2}(1+C)\right]\},
\end{equation}
and where $\hatz$ is the unit vector in the $z$-direction.
\label{eq:dim_nondim}
\end{subequations}
The incompressibility condition $\nabla\cdot\vecu=0$ is unchanged.

\section{Comparison with simulations: droplet impact on a surface}
\label{sec:impact}

In this section we generate data using numerical simulations of a 2D droplet impacting on a solid surface, at various values of the equilibrium contact angle.  In this way, we aim to build uup a database of simulation results to validate the correlations introduced in Section~\ref{sec:theory}.  We first of all describe the set-up of a test case, we then carefully analyse the convergence of the numerical results, and finally, we compare the simulation data to the correlations.  Throughout, we use the Diffuse Interface Method introduced in Section~\ref{sec:numerical}.

\subsection{Test Case}
\label{sec:test}

  To fix ideas, we focus on a particular case study, corresponding to a millimeter-scale water droplet falling in air, this is relevant to the industrial applications mentioned in the introduction.  The parameters of the test case are given in Table~\ref{tab:tab1}.
\begin{table}[htb]
	\centering
		\begin{tabular}{|c|c|c|}
		    \hline
			  &Water (L) & Air (G) \\
				\hline
			Dynamic Viscosity ($\mu$) &  $8.9\times 10^{-4}\,\mathrm{Pa}\,\mathrm{s}$ & $1.837\times 10^{-5}\,\mathrm{Pa}\,\mathrm{s}$\\
			Density ($\rho$)   &  $1000\,\mathrm{kg}\,\mathrm{m}^{-3}$  & $1.225\,\mathrm{kg}\,\mathrm{m}^{-3}$ \\
			\hline
		\end{tabular}
		
		\vskip 0.1in
		\begin{tabular}{|c|c|}
		\hline
			Droplet Radius ($R_0$) &  $3\,\mathrm{mm}$\\
			Surface Tension ($\surften$)   &  $0.072\,\mathrm{N}\,\mathrm{m}^{-1}$\\
			\hline
		\end{tabular}
		\caption{Physical parameters used for the test-case simulations.  A subscript $L$ is used to label water-based quantities and a subscript $G$ is used to label air-based quantities.}
		\label{tab:tab1}
\end{table}
We present our results in non-dimensional variables.  Lengthscales are made non-dimensional on the droplet radius $R_0$ and time is made dimensionless using the Bond timescale $T=R_0/U_0$, where $U_0=\sqrt{gR_0}$, where $g=9.81\,\mathrm{m}\,\mathrm{s}^{-2}$ is the acceleration due to gravity.  Densities and viscosities are made dimensionless on the liquid values.  Thus, in dimensionless variables, the acceleration due to gravity has unit strength.  Furthermore, the strength of the surface tension in Bond scaling is $1/\myBo$, where
\begin{equation}
\myBo=\frac{\rho_L g R_0^2}{\surften}
\end{equation}
In the same way, the strength of the viscosity in Bond scaling is $1/\myRe$, where
\begin{equation}
\myRe=\frac{\rho_L R_0 U_0}{\mu_L}=\frac{\rho_L g^{1/2} R_0^{3/2}}{\mu_L}.
\end{equation}
With the values in Table~\ref{tab:tab1}, we have $\myBo=1.226$ and $\myRe=578.0$. 
%
%
 Unless otherwise stated, in the rest of the work, results and discussion are presented in dimensionless variables.

\begin{figure}[htb]
	\centering
		\includegraphics[width=0.45\textwidth]{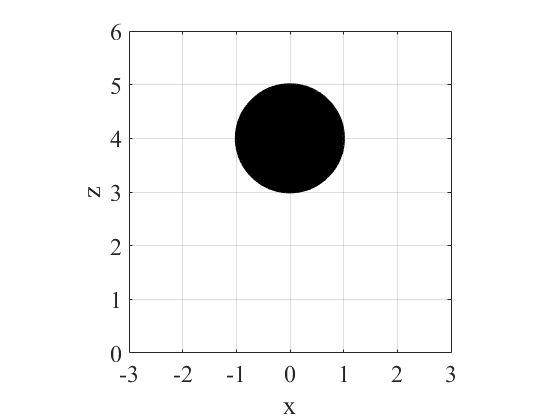}
		\caption{Initial condition of the droplet}
	\label{fig:velwprofile_init}
\end{figure}
We carry out simulations in a two-dimensional Cartesian geometry with the following domain
\begin{equation}
\Omega=\{(x,y)|-\tfrac{1}{2} L_x <x<\tfrac{1}{2}L_x, 0<z<L_z\},
\label{eq:omegadef}
\end{equation}
where the value  $L_z=6R_0$ is chosen.  Similarly, $L_x$ is chosen to be a multiple of $R_0$, with the exact value depending on the context. Gravity points in the negative $z$-direction.  Periodic boundary conditions on $\vecu$ are employed in the $x$-direction and no-slip boundary conditions on the walls at $z=0$ and $z=L_z$.    The simulations are initialized with zero velocity and zero pressure, and the droplet is released from rest with its centre initial located at $z=4R_0$ (e.g. Figure~\ref{fig:velwprofile_init}).  The pressure inside the droplet rapidly adjusts to take account of the Laplace pressure jump across the droplet interface.
%
%


\subsection{Benchmark Result and Convergence Study}
\label{sec:benchmark}

\begin{figure}[htb]
\centering
\subfigure[$\,\,t=2.6$]{\includegraphics[width=0.45\textwidth]{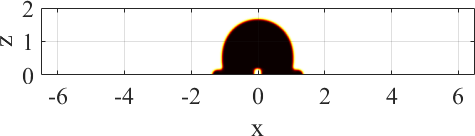}}
\subfigure[$\,\,t=2.8$]{\includegraphics[width=0.45\textwidth]{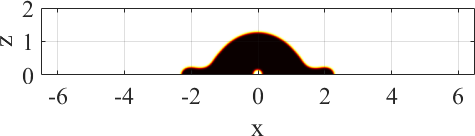}}
\subfigure[$\,\,t=3.0$]{\includegraphics[width=0.45\textwidth]{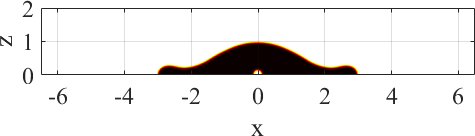}}
\subfigure[$\,\,t=3.6$]{\includegraphics[width=0.45\textwidth]{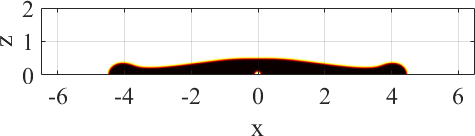}}
\subfigure[$\,\,t=4.0$]{\includegraphics[width=0.45\textwidth]{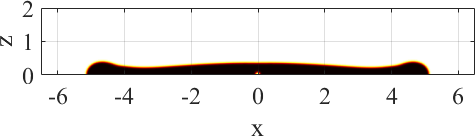}}
\subfigure[$\,\,t=6.1$]{\includegraphics[width=0.45\textwidth]{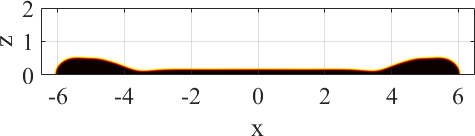}}
\caption{Sequence of states during the droplet spreading phase, $\thetaeq=90^\circ$}
	\label{fig:dimAdv}
\end{figure}

\begin{figure}[htb]
\centering
\subfigure[$\,\,t=8.3$]{\includegraphics[width=0.45\textwidth]{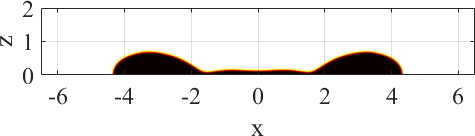}}
\subfigure[$\,\,t=9.0$]{\includegraphics[width=0.45\textwidth]{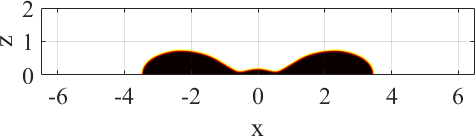}}
\subfigure[$\,\,t=9.7$]{\includegraphics[width=0.45\textwidth]{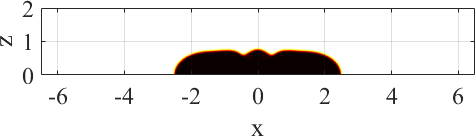}}
\subfigure[$\,\,t=10.3$]{\includegraphics[width=0.45\textwidth]{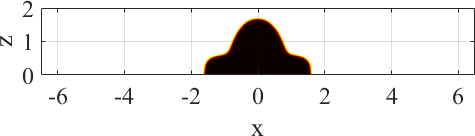}}
\caption{Sequence of states during the droplet receding phase, $\thetaeq=90^\circ$}
	\label{fig:dimRec}
\end{figure}

We first of all look at a case where the static contact angle $\thetaeq=90^{\circ}$, in this way, the boundary condition~\eqref{eq:dimGeom} becomes $\hatnwall\cdot\nabla C=0$ at $z=0$.  
Figure~\ref{fig:velwprofile_init} shows the initial droplet state.  Thereafter,
the droplet falls and impacts the substrate at dimensionless time $t \approx 2.5$ ($0.045\,\mathrm{s}$).  After the impact, the spreading phase takes place until maximum spreading is reached at dimensionless time $t \approx 6.0$. Figures~\ref{fig:dimAdv} and \ref{fig:dimRec} show this process in a Diffuse Interface simulation with static contact angle $\thetaeq=90^\circ$ and a mesh with 321 cells in the $z$-direction.
After the droplet has reached its maximum extent, the receding phase starts and the droplet contracts, as seen in Figure \ref{fig:dimRec}. At the end of the receding phase, the droplet adopts a compact shape (e.g. Figure~\ref{fig:dimRec}(e)), reminiscent of the moments just after the initial impact.   

\begin{figure}
	\centering
		\includegraphics[width=0.9\textwidth]{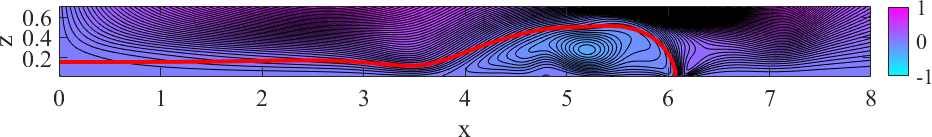}
		\caption{Instantaneous streamfunction at $t=6.1$, corresponding to Figure~\ref{fig:dimAdv}(f), near the point of maximum droplet spreading.  To present both axes on the same scale clearly, only the positive half-space $x>0$ is shown.}
	\label{fig:velpsiprofile_61}
\end{figure}

\begin{figure}
	\centering
		\includegraphics[width=0.9\textwidth]{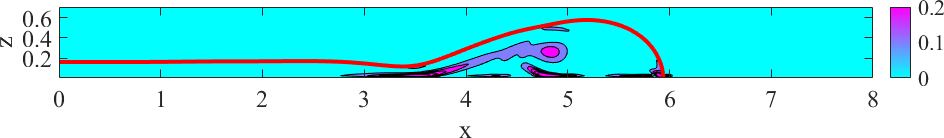}
		\caption{Instantaneous dissipation function $\text{Diss}=(\widehat{\mu}/\myRe)\left(u_x^2+u_z^2+w_x^2+w_z^2\right)$ at $t=6.1$, corresponding to Figure~\ref{fig:dimAdv}(f), near the point of maximum droplet spreading.  To present both axes on the same scale clearly, only the positive half-space $x>0$ is shown.  }
	\label{fig:dissprofile_61}
\end{figure}

The instantaneous streamfunction at $t=6.1$ is shown in Figure~\ref{fig:velpsiprofile_61}, near the point of maximum  spreading.  Near  $x=0$, the flow resembles that near a stagnation point, this corresponds to a region where the droplet resembles a thin sheet (the `lamella').   However, far from $x=0$, in the bulbous liquid region close to the droplet edge (the `rim'), there is a recirculation zone.
These flows generate substantial amounts of dissipation.  In Figure~\ref{fig:dissprofile_61} we plot the dissipation function $\text{Diss}=(\widehat{\mu}/\myRe)\left(u_x^2+u_z^2+w_x^2+w_z^2\right)$, again at $t=6.1$.  There is a large contribution to the dissipation due to the boundary layer in the droplet lamella.  A second large contribution inside the droplet rim is also in evidence in the figure, this can be associated with a boundary layer which forms inside the rim itself, this arises due to the flow which circulates inside the droplet rim.  There is a third seemingly large contribution to the dissipation at the contact line itself.  However, this contribution occurs over a very small region (around 20 gridpoints in each direction).  The total dissipation $\int_\Omega (\text{Diss})\mathd^2 x$ is measured to be $0.191$, the total dissipation without this small region included is measured to be $0.189$, a $1\%$ difference.  Thus, this third contribution to the total dissipation is not very important.

\begin{figure}[htb]
	\centering
	\includegraphics[width=0.8\textwidth]{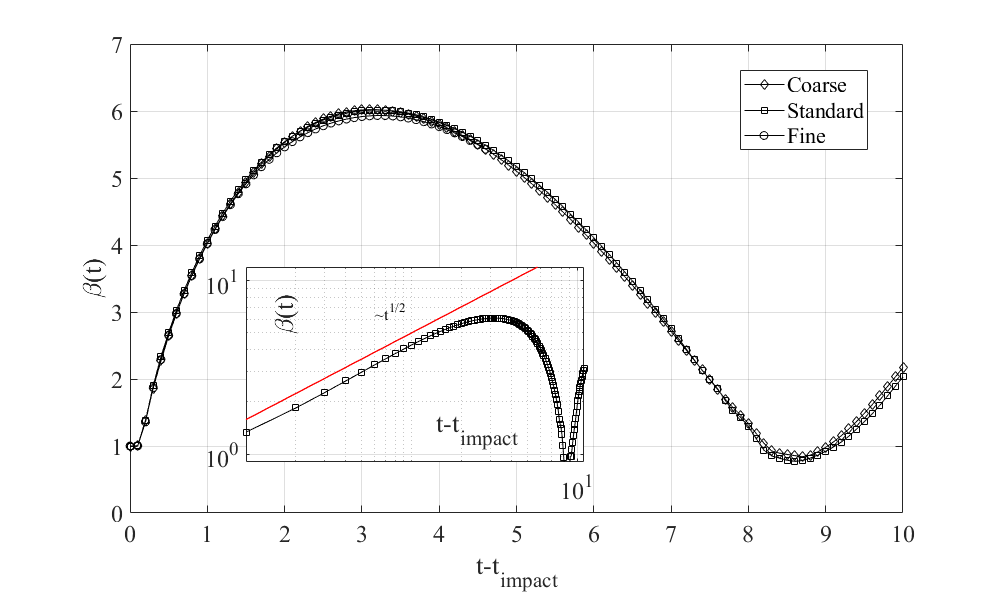}
	\caption{Spreading ratio of the droplet showing the attainment of grid-independent results.  Inset -- the same as the main figure, only using a log-log scale, and with $t-t_{impact}$ on the horizontal axis.  Here $t_{impact}$ refers to the time of droplet impact.}
	\label{fig:dimMesh}
\end{figure}

We also show the convergence of the numerical simulations with respect to grid refinement.   Hence, different uniform meshes are prepared (with $\Delta x=\Delta z$), each characterized by the number of cells in the vertical direction (denoted by $N$).  Following standard practice~\cite{yue2011can}, the Cahn number varies as $\myCn\propto \Delta x$. The parameters used in the convergence study are shown in Table \ref{tab:convergence1}. 
%
\begin{table}[htb]
	\centering
	\begin{tabular}{|c|c|c|c|c|c|}
		\hline
		$N$ &  $\Delta t$ & $\Delta x$ & $\myCn$ & $\mydiff$ & Label\\
		\hline 
		\hline
		%
		161 &   $10^{-4}$         &0.0375    & $(4/3)\Delta x$ & $\Delta x$ & Coarse \\
		321 &   $10^{-4} $        &0.01875    & \dittotikz   & \dittotikz &Standard \\
		641 &    $10^{-4}/2$      &0.009375   &  \dittotikz & \dittotikz &Fine\\
		\hline
	\end{tabular}
	\caption[Parameters of the convergence study in DIM]{Convergence study for the Diffuse Interface Method, for $\thetaeq=90^{\circ}$. To ensure a CFL number less than one, the timestep is halved for the  $N=641$ case.}
	\label{tab:convergence1}
\end{table}
The results are shown in Figure~\ref{fig:dimMesh}, where the spreading ratio $\beta(t)$ of the droplet is plotted against time. This ratio is calculated as $\beta(t)=r(t)/R_0$, 
where $r(t)$ is the maximum extent of the droplet as it spreads across the substrate, at time $t$, and $R_0$ is the initial radius of the droplet. Clearly, convergence of the numerical method is achieved on all meshes used.  For the remainder of the work, the `standard' mesh with $321$ gridpoints in the $z$-direction is used. 

From Figure~\ref{fig:dimMesh}, it is also seen that the initial phase of the droplet spreading is characterized by the relation $\beta(t)\propto (t-t_{impact})^{1/2}$, where $t_{impact}$ is the time of the initial droplet impact.  This scaling is shown in the inset of Figure~\ref{fig:dimMesh}, and is  is consistent with a balance between the capillary pressure and the inertial pressure at the contact line~\cite{legendre2015comparison}, valid just after droplet impact.  Specifically, the balance is $\rho_L (\mathd r/\mathd t)^2\propto \surften R_0/r^2$.
Finally, a key conclusion from Figure~\ref{fig:dimMesh} is that the simulations are independent of mesh size for the parameters chosen in Table~\ref{tab:convergence1}.  Therefore, throughout the remainder of the work, the standard mesh
resolution is used.

\begin{figure}
	\centering
		\includegraphics[width=0.8\textwidth]{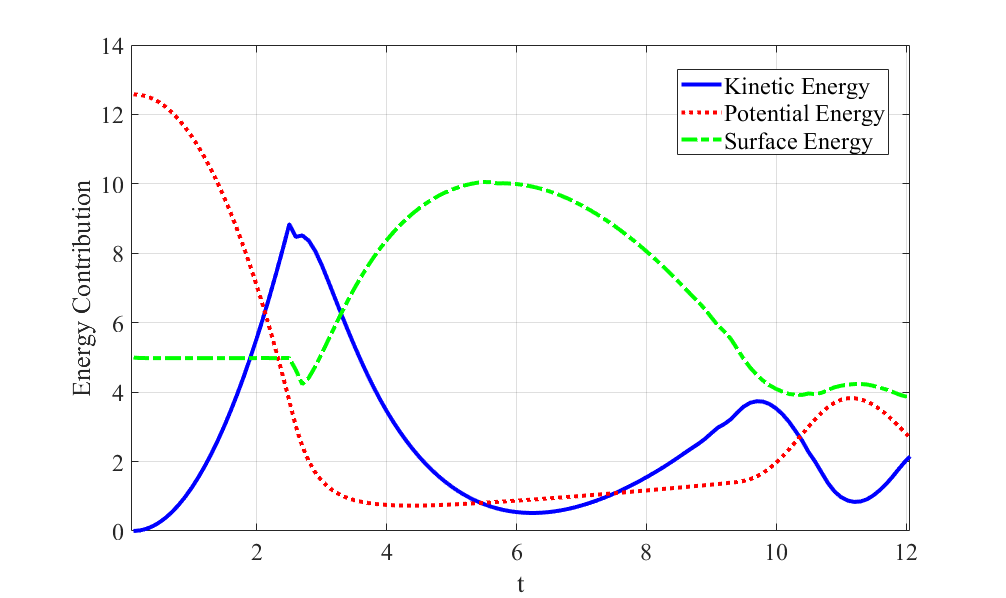}
		\caption{Energy budget for the case $\thetaeq=90^\circ$.}
	\label{fig:energy_budget90}
\end{figure}

To further understand the droplet-spreading dynamics up to the point of maximum spread, we look at the total energy of the droplet, made up of kinetic energy, potential energy and surface energy:
\[
E_D=\int_\Omega \left(\tfrac{1}{2}\vecu^2+ z\right) \left[\tfrac{1}{2}(1+C)\right]\mathd^3x+F[C].
\]
As we are working in a two-dimensional / three-dimensional cylindrical droplet configuration, the relevant dimensions are $x$ and $z$, the third dimension into the plane of the page can be ignored.
The total system energy  involves further contributions from the gas-side kinetic and potential energy. However, since $\rho_G\ll \rho_G$, these contribution is negligible, and hence, the droplet energy $E_D$ is a good approximation of the total system energy.  Thus, `total energy' and `droplet energy' can be used interchangeably in what follows.
The energy `budget' showing the different contributions to the droplet energy is shown in Figure~\ref{fig:energy_budget90}.
Overall, the main contribution to the droplet energy after impact is the surface energy.  Furthermore, surface energy is maximum when kinetic energy is minimum, this coincides with maximum droplet spread. 

\begin{figure}
	\centering
		\subfigure[]{\includegraphics[width=0.45\textwidth]{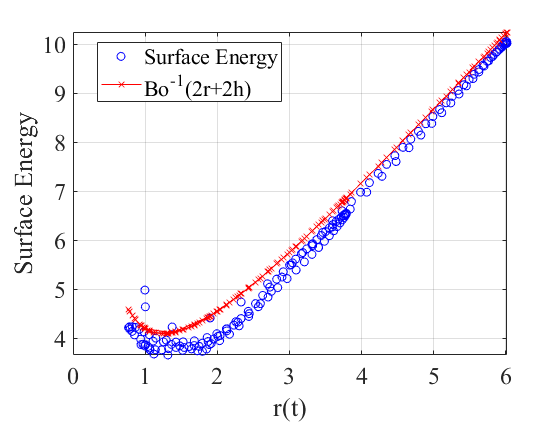}}
		\subfigure[]{\includegraphics[width=0.45\textwidth]{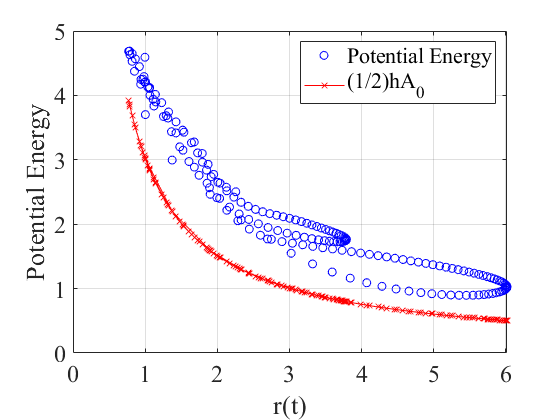}}
		\caption{Dependency of the different energy terms on the instantaneous droplet extent $r(t)$}
	\label{fig:scaling90}
\end{figure}

The dependency of the surface energy and potential energy on the instantaneous droplet radius $r(t)$ is shown in Figure~\ref{fig:scaling90}. 
The surface energy can be very well approximated by treating the droplet as a two-dimensional rectangular shape -- the analogue for three-dimensional axisymmetric droplets would be a disk,  this is a common approach in the literature~\cite{chandra1991collision}. In this context, the droplet extent is $r$ and the droplet height is $h=A_0/(2r)$, where $A_0=\pi R_0^2$ is the initial droplet area (volume).  In this way, the approximate surface energy is $(1/\myBo)(2r+2h)$, Figure~\ref{fig:scaling90}(a) shows excellent agreement between this rather crude approximation and the results of the numerical simulations.  Furthermore, the gravitational potential energy is well captured by the same rectangular-shape approximation, with the gravitational potential energy estimated as $ (h/2)A_0$ (Figure~\ref{fig:scaling90}(b)) -- the factor of half is needed as this corresponds to the elevation of the droplet's center of mass.     We  use these findings in what follows to validate our correlations for $\betamax$.

\subsection{Correlations for the Spreading Radius}

We carry over the correlations developed in Section~\ref{sec:theory} to the present case.  As we are dealing with a smooth surface with no contact-angle hysteresis, we use the equilibrium contact angle in place of the advancing one.  Also, as we are dealing with Bond-number scaling, some translation of the previous results is required.  In particular, gravitational potential energy is now required in the energy balance.  The elevation of the droplet centre of mass is approximated by $h/2=A_0/(4r_{max})$.  In this way, the energy budget~\eqref{eq:eb1} becomes:
\begin{multline}
\rho_L g H_0 (\pi R_0^2)+2\pi R_0\surften =
\surften \left(\Pi-2r_{max}\cos\thetaeq\right) +\rho_L g (\pi R_0^2) d_c\\
+\frac{a}{\sqrt{\myRe}}\rho_L U_0^2 \beta_{max}\sqrt{\beta_{max}-1}\,R_0^2
+b \rho_L gH_0 (\pi R_0^2).
\label{eq:energyGRAV}
\end{multline}
Divide across by $\rho_L g R_0^3\equiv\rho_L U_0^2 R_0^2$ to obtain:
\begin{multline}
\pi(H_0/R_0)+\frac{2\pi}{\myBo}  =
\frac{1}{\myBo}\left[2\beta_{max}\left(1-\cos\thetaeq\right)+ \frac{\pi}{\beta_{max}}\right]+\pi\left(\frac{\pi}{4\beta_{max}}\right)\\
+\frac{a}{\sqrt{\myRe}}\beta_{max}\sqrt{\beta_{max}-1}+\pi(H_0/R_0)b.
\label{eq:correlationGRAV}
\end{multline}

\begin{figure}[htb]
	\centering
		\includegraphics[width=0.5\textwidth]{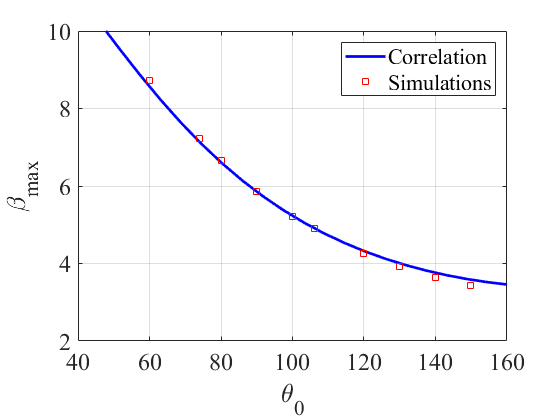}
		\caption{Dependence of $\beta_{max}$ on the static contact angle for fixed $\myBo$ and $\myRe$.  Squares: Simulations.  Solid line: the correlation~\eqref{eq:correlation}.}
	\label{fig:prediction}
\end{figure}

We apply the correlation in Equation~\eqref{eq:correlationGRAV} to Figure~\ref{fig:prediction}.     The parameters $a$ and $b$ are fitted \textit{only once} to the case $\thetaeq=90^\circ$, to minimize the difference between $\beta_{max}$ from the simulations and $\beta_{max}$ from the correlation.  Thereafter, the values of $a$ and $b$ are fixed and the correlation~\eqref{eq:correlation} is used predictively, to determine the dependency of $\beta_{max}$ on $\thetaeq$ at fixed $\myBo$ and $\myRe$.   The results of the prediction are shown in Figure~\ref{fig:prediction}.  The values of $a$ and $b$ are given in Table~\ref{tab:abc} (the `two-parameter model' therein).
\begin{table}[htb]
	\centering
		\begin{tabular}{|c|c|c|c|}
		\hline
		                    & $a$ & $b$ & $c$ \\
												\hline
		Two-parameter model & 3.2096 & 0.3056 & $-$\\
		Three-parameter model & 2.4963 &  0.5115 &  0.2738 \\
		\hline
		\end{tabular}
		\caption{Estimated values of the fitting parameters for the models used}
		\label{tab:abc}
\end{table}

\begin{figure}[htb]
	\centering
		\subfigure[]{\includegraphics[width=0.45\textwidth]{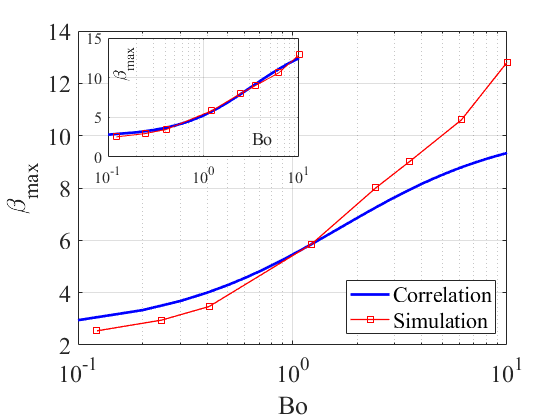}}
		\subfigure[]{\includegraphics[width=0.45\textwidth]{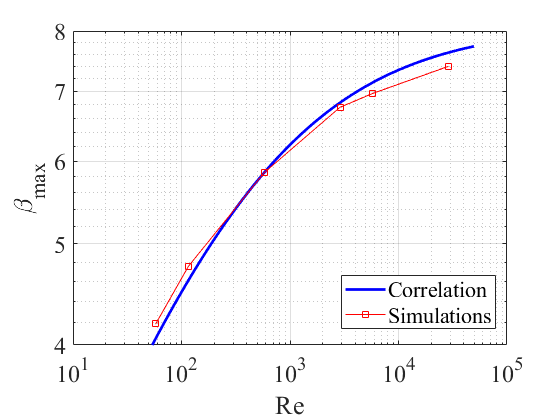}}
		\caption{Dependence of $\beta_{max}$ on the parameters $\myBo$ and $\myRe$ for fixed $\thetaeq=90^\circ$. 
		Panel (a): Fixed $\myRe=578$ and varying $\myBo$; Panel (b): fixed $\myBo=1.225$ and varying $\myRe$.
		Squares: Simulations.  Solid line: the correlation~\eqref{eq:correlation} (`two-parameter model').  The inset in Panel (a) shows the three-parameter model.}
	\label{fig:prediction1}
\end{figure}

To validate the correlation~\eqref{eq:correlationGRAV} further, we have repeated the simulation described previously (Section~\ref{sec:benchmark}) with $\thetaeq=90^\circ$, but with various values of the surface tension $\surften$ and liquid viscosity $\mu_L$.  In this way we can investigate the effect of the Bond number and the Reynolds number of the maximum droplet spread.   The results are shown in Figure~\ref{fig:prediction1}. 
Overall, there is good agreement between the correlation~\eqref{eq:correlation} and the simulation data over  three orders of magnitude of $\myBo$ (Panel (a)).  There is excellent agreement between the correlation and the simulation data over several orders of magnitude of $\myRe$ (Panel (b)).  This last result lends weight to the theory that the boundary layer is a main source of dissipation in the problem, with the boundary-layer dissipation term scaling as $\myRe^{-1/2}$.

\begin{figure}
	\centering
		\includegraphics[width=0.9\textwidth]{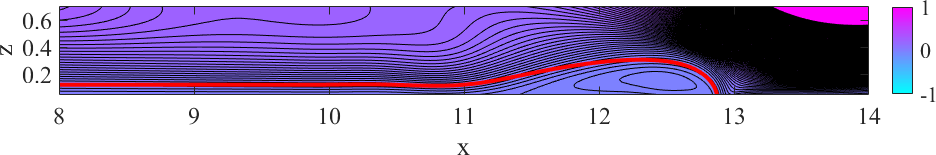}
		\includegraphics[width=0.9\textwidth]{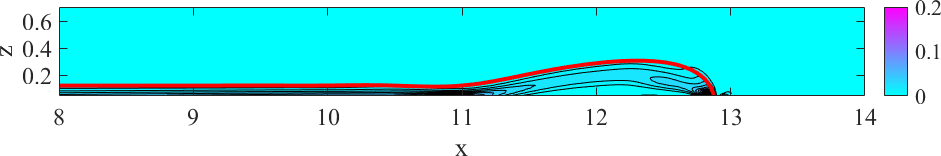}
		\caption{Results at $\myBo=10$ and $\myRe=578$.  (a) Instantaneous streamfunction at $t=10.3$,  near the point of maximum droplet spreading.  (b)  Instantaneous dissipation at the same time.  To present both axes on the same scale clearly, only the range $8\leq x\leq 14$ is shown.  The symmetry point is at $x=0$.}
	\label{fig:largeBo}
\end{figure}

To improve the agreement between the simulation data and the theory in the case where $\myBo$ is varied, we have investigated further the flow pattern and dissipation pattern inside the droplet in the case of very high $\myBo$ (Figure~\ref{fig:largeBo}).  
%
These results may be compared with Figure~\ref{fig:velpsiprofile_61} (streamfunction), 
and Figure~\ref{fig:dissprofile_61} (dissipation).  For the purpose of comparison, the same scales on the colour bar are used throughout.  
The recirculation zone at high $\myBo$ is more organized in Figure~\ref{fig:largeBo}(a), compared to Figure~\ref{fig:velpsiprofile_61}.  Also, the gradient in the streamfunction  is smaller in magnitude than before, this can be seen by the sharp color gradient in the recirculation zone in Figure~\ref{fig:velpsiprofile_61}, as opposed to the complete absence of such a color gradient in the recirculation zone in Figure~\ref{fig:largeBo}(a).  
All of these makes for a reduced level of dissipation, this can be seen by comparing Figure~\ref{fig:dissprofile_61} with Figure~\ref{fig:largeBo}(b).  Although boundary-layer dissipation is still in evidence in  Figure~\ref{fig:largeBo}(b), overall, the amount of dissipation is less than before.  This suggests that the a reduction in the `head loss' dissipation term is appropriate at large values of $\myBo$.

These differences call for a more detailed approach to the modeling of the droplet rim, which provides scope for future work.  In the meantime, we model the change in the flow regime using a fitted model, $\Delta E_h=\left[b/(1+c\myBo)\right]\rho_L g H_0 (\pi R_0^2\lambda)$, where $b$ and $c$ are coefficients.  As before, the model coefficients are obtained via nonlinear least-squares fitting (now $a$, $b$, and $c$).  The nonlinear optimization is done using constraints, such that $b\leq 1$ and $c\geq 0$.  These constraints are required for conservation of energy.  The results of this updated model are shown 
in the inset of Panel (a) in Figure~\ref{fig:prediction1}.  Estimated parameter values are given in  Table~\ref{tab:abc} (the `three-parameter model' therein).

\section{Comparison with simulations and experiments: head-on collision of two droplets}
\label{sec:head-on}

In this section we consider the maximum spreading radius in case of a head-on collision of two droplets.
We examine the maximum spreading radius as obtained in simulations that already exist in the literature, and supplement these with our own. We investigate the extent to which the simple correlation~\eqref{eq:inviscid1} for head-on collisions fits the numerical data.  We also investigate whether the same simple correlation applies to experiments on the collision of quasi-2D cylindrical structures.

\subsection{Comparison with Simulations}

\begin{figure}[htb]
	\centering
		\includegraphics[width=0.45\textwidth]{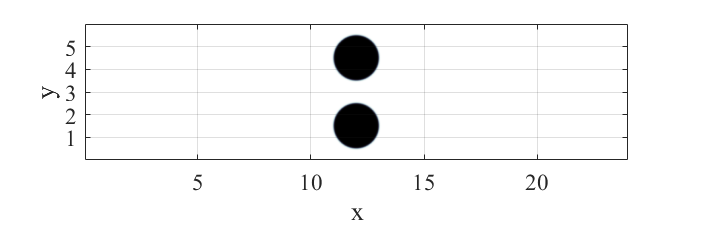}
		\caption{Initial condition for the head-on collision.  The upper droplet has a velocity $w=-U_0$ and the lower droplet has a velocity $w=+U_0$.}
	\label{fig:IC_headon}
\end{figure}

We start by generating data using numerical simulations of a head-on collision between two 2D droplets, with a view to building  up a database of simulation results to validate the correlation~\eqref{eq:inviscid1}.  We further compare our results with earlier results on head-on collisions by Wu et al.~\cite{wu2021decoupled}  For these purposes, we again use the diffuse-interface method and the numerical parameters in Section~\ref{sec:impact}, with the initial condition shown in Figure~\ref{fig:IC_headon}.  Gravity is set to zero. The key dimensionless variables are thus $\myWe=\rho_L U_0^2 R_0/\sigma$ and $\myRe=\rho_L U_0 R_0/\mu_L$. 

\begin{figure}[htb]
	\centering
		\subfigure[$\,\,t=0.6$]{\includegraphics[width=0.45\textwidth]{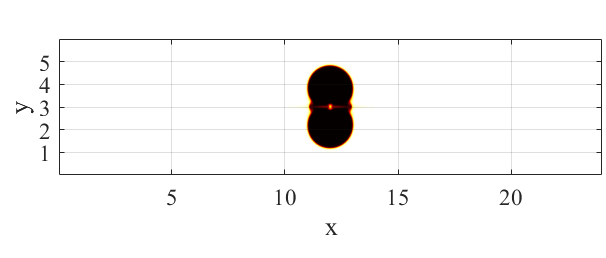}}
		\subfigure[$\,\,t=2.0$]{\includegraphics[width=0.45\textwidth]{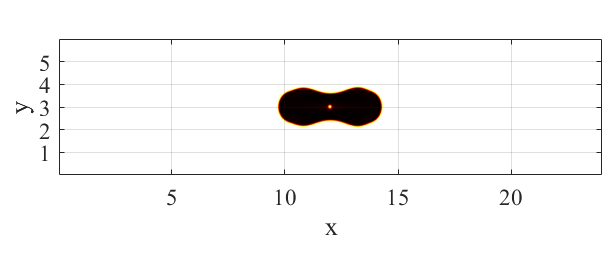}}
		\subfigure[$\,\,t=3.0$]{\includegraphics[width=0.45\textwidth]{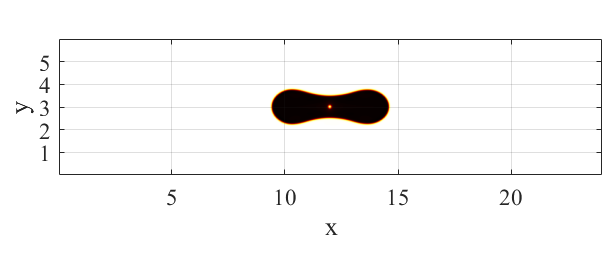}}
		\subfigure[$\,\,t=4.0$]{\includegraphics[width=0.45\textwidth]{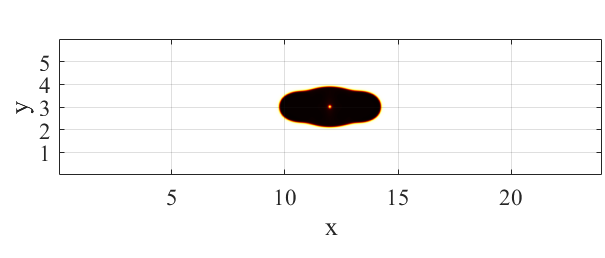}}
		\caption{Sequence of states for the head-on collision.  $\myWe=1.225$ and $\myRe=578$.  Maximum spreading occurs at $t\approx 3.2$.}
	\label{fig:headon}
\end{figure}

\begin{figure}[htb]
	\centering
		\subfigure[$\,\,t=2.0$]{\includegraphics[width=0.45\textwidth]{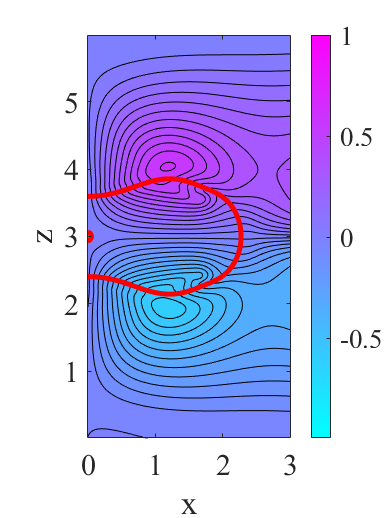}}
		\subfigure[$\,\,t=3.0$]{\includegraphics[width=0.45\textwidth]{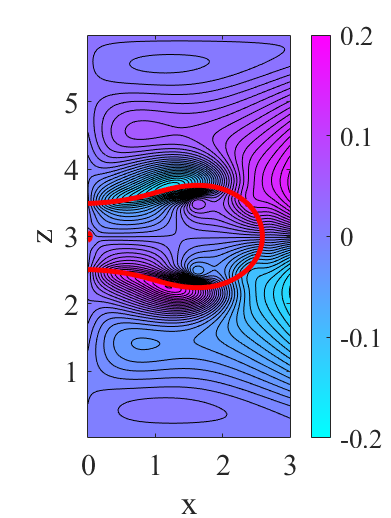}}
		\caption{Snapshots of streamfunction in case of head-on collision.  $\myWe=1.225$ and $\myRe=578$. }
	\label{fig:headon1}
\end{figure}

\begin{figure}
	\centering
		\includegraphics[width=0.6\textwidth]{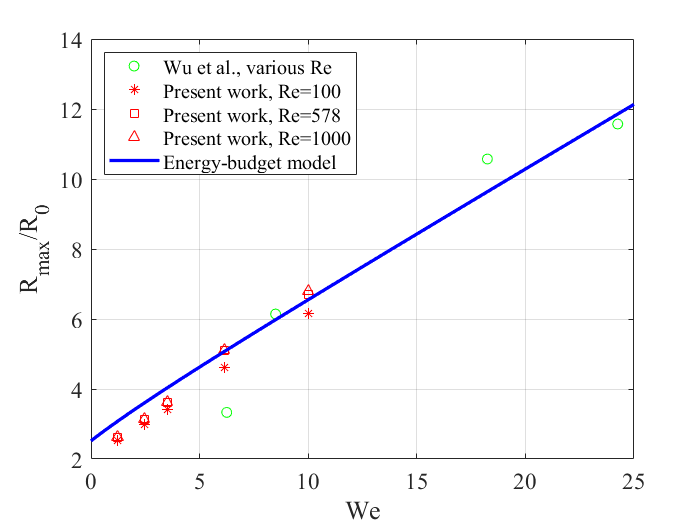}
		\caption{Comparison between the model~\eqref{eq:inviscid1} and the simulations by Wu et al.~\cite{wu2021decoupled}}
	\label{fig:wu_head_on}
\end{figure}

Sample results  are shown in Figure~\ref{fig:headon}.  The results show droplet impact, spreading, and retraction, similar to what was observed in Section~\ref{sec:impact} in case of droplet impact on a solid substrate.  Snapshots of the streamfunction are shown in Figure~\ref{fig:headon1}.  These show a recirculation zone in the rim, which gives rise to head loss.
%
%
%
%
%
The maximum spreading radius as a function of $\myWe$ and $\myRe$ is plotted in Figure~\ref{fig:wu_head_on}.  Here, we also include the maximum spreading radius computed from the simulations by Wu et al.~\cite{wu2021decoupled}
In this same figure, we also compare the results of the numerical simulations to the correlation~\eqref{eq:inviscid1}.   For this reason, we fit the data from the paper of Wu et al.~\cite{wu2021decoupled} to the correlation, for which non-linear least-squares fitting yields $b=0.5355$.  The data involve the collation of results at various Reynolds numbers, on the understanding that the Weber-number effect is dominant in determining $\beta_{max}$. 
We use only those cases from the reference that do not involve film rupture, as such rupture in 2D is a numerical artefact.  The results show that the energy loss can be well captured across the range of Weber numbers considered by a simple `head loss' factor $b$, independent of Reynolds number.  Crucially, the neglect of the head loss leads to an over-prediction of the maximum spreading radius.  Our own numerical results fall on the same curve, albeit that the slope of the trend line exhibits a mild Reynolds-number dependence at $\myRe=100$.  

\subsection{Comparison with Experiments}

We also look at the performance of the model with respect to experimental data.  N\'eel et al.~\cite{neel2020fines} generate two toroidal liquid cylinders on the $\mathrm{mm}$ scale and engineer their head-on collision.  Upon impact, a portion of the tori elongate to form cylindrical structures.  The cylinders impact and spread in a direction perpendicular to the plane of the tori.  The authors argue for the impact to be treated as the head-on collision of two cylindrical liquid structures.   As such, the two-dimensional droplet-impact problem studied herein has an  experimental analogue.  Upon impact, the two liquid cylinders form a fused lamella which elongates until a maximum length is attained.   Beyond a critical Weber number, the lamella breaks up under a Rayleigh--Taylor instability in a process resembling droplet splash in the radially symmetric case.  
\begin{figure}
	\centering
		\includegraphics[width=0.6\textwidth]{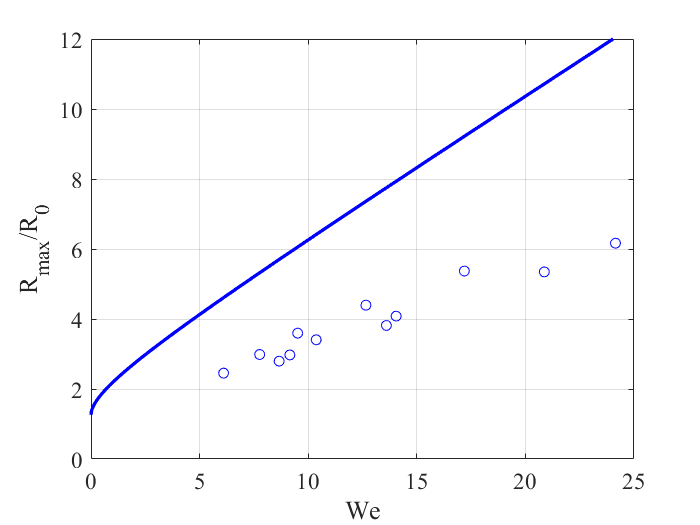}
		\caption{Comparison between the model~\eqref{eq:inviscid1} and the experiments by N\'eel et al.~\cite{neel2020fines}, in case of impact of two tori.  The Weber number on the $x$-axis is $\myWe=\rho_L U_0^2 R_0/\sigma$; the Weber number used in the reference is $8\rho_L U_0^2R_0/\sigma$.  The Reynolds number $\rho_L U_0 R_0/\mu_L$ is $O(10^5)-O(10^6)$.}
	\label{fig:neel_head_on}
\end{figure}

We compare the experimental results by N\'eel et al.~\cite{neel2020fines} with the model in Figure~\ref{fig:neel_head_on}, using $b=0.5355$ as before.  
As such, we make no attempt to fit the correlation to the data, and use only those model parameters determined from the previous simulation results.  The correlation reproduces the trend in the data but the quantitative agreement is poor.  Expecting  a simple 2D model to provide exact quantitative agreement in case of a 3D toroidal impact problem may be unrealistic.  The agreement may be improved by explicitly accounting for spontaneous surface dissipation in the energy budget (thus lowering the $y$-intercept of the curve in Figure~\ref{fig:neel_head_on}), and then refitting the model to the data.  We do not pursue this approach here.  Crucially however, our correlation does exhibit the observed scaling behaviour $\lam_{max}\sim \myWe$ at large Weber number.  Such scaling is a signature of onset of 2D behaviour in the spreading problem (3D spreading involves $\lam_{max}\sim \myWe^{1/2}$ at large Weber number).

Finally, we  explore the overlap between the present energy-budget analysis and the simple one-dimensional mechanistic model for the lamella extension $\betamax$ put forward by N\'eel et al.~\cite{neel2020fines}  The simple model in the reference predicts $\betamax=1+(\pi/4)\myWe$.  The slope here the same as in Equation~\eqref{eq:inviscid1}, with $b=0$.  This is a drastic over-prediction of $\betamax$; hence, the head-loss factor is necessary to provide agreement between the experiments and the models.


\section{Discussion and Conclusions}
\label{sec:conc}

Summarizing, we have looked at the phenomenon of droplet spreading upon impact in 2D, in case of impact on a solid substrate, as well as the head-on collision of two droplets.  The setup we consider is a rather unusual physical scenario.  However, it is studied in the computational literature as a testbed for similar but more computationally expensive simulations in three dimensions.  Furthermore, recent experiments on liquid sheets suggest that such a setup may serve as a useful analogy for real 3D phenomena.
%
%
%
The main result is an energy-budget analysis, which produces a correlation for the maximum spreading radius as a function of $\myWe$ and $\myRe$.  To match the experiments and the simulations results, the correct modeling of dissipation is crucial.  

\paragraph*{Droplet impact on a solid substrate:}  The data suggest at least two key dissipation channels.  Boundary-layer dissipation in the lamella  is important: without it, the model would greatly over-estimate the maximum droplet spread, $r_{max}$.  Our simulations reveal clearly the boundary layer, both in the instantaneous streamfunction, and the instantaneous dissipation function.   

A second source of dissipation  is the so-called `head loss'.  Classically, `head loss' (e.g. in a pipe flow) refers to the loss of pressure head due to wall friction (major head loss), or to loss of pressure due to sharp changes in the pipe geometry, which lead to recirculation zones (minor head loss).  In the context of droplet spread, head loss is taken to mean friction losses due to all loss channels other than the main boundary-layer dissipation. In this work, we see that such loss channels include  a recirculation flow in the rim.  

The simulations show that the importance of head loss to the overall energy budget depends on surface tension: the lower the surface tension, the less important the head loss becomes.  As surface tension is a conservative force, the change in the amount of head loss that occurs when the surface tension is lowered must be through an indirect route.  
%
%
Consequently, the `head loss', previously described as a simple (universal) fraction of the initial kinetic energy, is more complicated.  Our three-parameter correlation, based again on an energy budget, fits the simulation data very well, over several orders of magnitude of $\myWe$ and $\myRe$.  


\paragraph*{Droplet impact on a solid substrate:} In case of head-on collision of two droplets, energy dissipation is required to make the energy-budget analysis agree with the experimental and numerical results.  Although the simulations show that no boundary layer is present, they do show a significant region of dissipation near the droplet `triple point' just after impact.  A simple `head loss' argument can again be applied to the energy-budget analysis, whereby the dissipated energy is equated with a fraction of the initial kinetic energy.  This approach gives good agreement between the theoretical prediction for the spreading radius, and the numerical simulations, at sufficiently large Reynolds number.

Overall, the correlations introduced here agree well with the simulation data, over several orders of magnitude of $\myWe$ and $\myRe$.  However, the argument based on head loss, while based on an earlier work on energy-budget analysis for droplet spreading~\cite{wildeman2016spreading}, still involves parameters which have to be fitted to the data.  A more physics-informed approach (such as a rim-lamella model, tailored to two dimensions) may be of use to estimate these parameters.  As such, the present work -- including the comprehensive database of numerical simulation results -- may serve as a basis for future work for a more physics-informed approach to the head-loss calculation, involving a more detailed description of the rim-lamella dynamics.


\subsection*{Acknowledgments}

The authors  acknowledge the Irish Centre for High-End Computing (ICHEC) for the provision of computational facilities and support (project ID: ndphy119c).
LON has also been supported by the ThermaSMART network. The ThermaSMART network has received funding from the European Union’s Horizon 2020 research and innovation programme under the Marie Sklodowska–Curie grant agreement No. 778104.




\begin{thebibliography}{10}

\bibitem{yarin2006drop}
Alexander~L Yarin.
\newblock Drop impact dynamics: splashing, spreading, receding, bouncing….
\newblock {\em Annu. Rev. Fluid Mech.}, 38:159--192, 2006.

\bibitem{valluri2015}
P. J. Sáenz, K.~Sefiane, J.~Kim, O. K. Matar, and P.~Valluri.
\newblock Evaporation of sessile drops: a three-dimensional approach.
\newblock {\em Journal of Fluid Mechanics}, 772:705–739, 2015.

\bibitem{Moghtadernejad2020}
Sara Moghtadernejad, Christian Lee, and Mehdi Jadidi.
\newblock An introduction of droplet impact dynamics to engineering students.
\newblock {\em Fluids}, 5(3), 2020.

\bibitem{josserand2016drop}
Christophe Josserand and Sigurdur~T Thoroddsen.
\newblock Drop impact on a solid surface.
\newblock {\em Annual review of fluid mechanics}, 48:365--391, 2016.

\bibitem{lejeune2018edge}
Sophie Lejeune, Tristan Gilet, and Lydia Bourouiba.
\newblock Edge effect: liquid sheet and droplets formed by drop impact close to
  an edge.
\newblock {\em Physical Review Fluids}, 3(8):083601, 2018.

\bibitem{neel2020fines}
B~N{\'e}el, Henri Lhuissier, and E~Villermaux.
\newblock ‘fines’ from the collision of liquid rims.
\newblock {\em Journal of Fluid Mechanics}, 893:A16, 2020.

\bibitem{ding2007diffuse}
Hang Ding, Peter~DM Spelt, and Chang Shu.
\newblock Diffuse interface model for incompressible two-phase flows with large
  density ratios.
\newblock {\em Journal of Computational Physics}, 226(2):2078--2095, 2007.

\bibitem{shin2009simulation}
Seungwon Shin and Damir Juric.
\newblock Simulation of droplet impact on a solid surface using the level
  contour reconstruction method.
\newblock {\em Journal of mechanical science and technology}, 23:2434--2443,
  2009.

\bibitem{gupta2011two}
Amit Gupta and Ranganathan Kumar.
\newblock Two-dimensional lattice boltzmann model for droplet impingement and
  breakup in low density ratio liquids.
\newblock {\em Communications in Computational Physics}, 10(3):767--784, 2011.

\bibitem{wu2017dynamics}
Zhenlong Wu and Yihua Cao.
\newblock Dynamics of initial drop splashing on a dry smooth surface.
\newblock {\em PLoS One}, 12(5):e0177390, 2017.

\bibitem{wu2021decoupled}
Yongyong Wu, Nan Gui, Xingtuan Yang, Jiyuan Tu, and Shengyao Jiang.
\newblock A decoupled and stabilized lattice boltzmann method for multiphase
  flow with large density ratio at high reynolds and weber numbers.
\newblock {\em Journal of Computational Physics}, 426:109933, 2021.

\bibitem{rafi2022two}
Abid~Hasan Rafi, Mohammad~Rejaul Haque, and Dewan~Hasan Ahmed.
\newblock Two-dimensional analogies to the deformation characteristics of a
  falling droplet and its collision.
\newblock {\em Archive of Mechanical Engineering}, pages 21--43, 2022.

\bibitem{wildeman2016spreading}
Sander Wildeman, Claas~Willem Visser, Chao Sun, and Detlef Lohse.
\newblock On the spreading of impacting drops.
\newblock {\em Journal of fluid mechanics}, 805:636--655, 2016.

\bibitem{eggers2010drop}
Jens Eggers, Marco~A Fontelos, Christophe Josserand, and St{\'e}phane Zaleski.
\newblock Drop dynamics after impact on a solid wall: theory and simulations.
\newblock {\em Physics of fluids}, 22(6):062101, 2010.

\bibitem{roisman2002normal}
Ilia~V Roisman, Romain Rioboo, and Cameron Tropea.
\newblock Normal impact of a liquid drop on a dry surface: model for spreading
  and receding.
\newblock {\em Proceedings of the Royal Society of London. Series A:
  Mathematical, Physical and Engineering Sciences}, 458(2022):1411--1430, 2002.

\bibitem{roisman2009inertiaP2}
Ilia~V Roisman.
\newblock Inertia dominated drop collisions. ii. an analytical solution of the
  navier--stokes equations for a spreading viscous film.
\newblock {\em Physics of Fluids}, 21(5):052104, 2009.

\bibitem{chandra1991collision}
S~Chandra and CT~Avedisian.
\newblock On the collision of a droplet with a solid surface.
\newblock {\em Proceedings of the Royal Society of London. Series A:
  Mathematical and Physical Sciences}, 432(1884):13--41, 1991.

\bibitem{villermaux2011drop}
Emmanuel Villermaux and Benjamin Bossa.
\newblock Drop fragmentation on impact.
\newblock {\em Journal of Fluid Mechanics}, 668:412--435, 2011.

\bibitem{park2003single}
Heungsup Park, Wallace~W Carr, Junyong Zhu, and Jeffrey~F Morris.
\newblock Single drop impaction on a solid surface.
\newblock {\em AIChE journal}, 49(10):2461--2471, 2003.

\bibitem{huang2018energetic}
Hai-Meng Huang and Xiao-Peng Chen.
\newblock Energetic analysis of drop’s maximum spreading on solid surface
  with low impact speed.
\newblock {\em Physics of Fluids}, 30(2), 2018.

\bibitem{willis2003binary}
Keeney Willis and Melissa Orme.
\newblock Binary droplet collisions in a vacuum environment: an experimental
  investigation of the role of viscosity.
\newblock {\em Experiments in fluids}, 34(1):28--41, 2003.

\bibitem{ewetola2021control}
Michael Ewetola, Rodrigo Ledesma-Aguilar, and Marc Pradas.
\newblock Control of droplet evaporation on smooth chemical patterns.
\newblock {\em Physical Review Fluids}, 6(3):033904, 2021.

\bibitem{naraigh2014linear}
Lennon {\'O N\'araigh}, Prashant Valluri, David~M Scott, Iain Bethune, and
  Peter~DM Spelt.
\newblock Linear instability, nonlinear instability, and ligament dynamics in
  three-dimensional laminar two-layer liquid/liquid flows.
\newblock {\em Journal of fluid mechanics}, 750:464--506, 2014.

\bibitem{naraigh2015flow}
Lennon {\'O N\'araigh}, Selma Shun, and Aurore Naso.
\newblock Flow-parametric regulation of shear-driven phase separation in two
  and three dimensions.
\newblock {\em Physical Review E}, 91(6):062127, 2015.

\bibitem{fannon2016high}
James Fannon, Jean-Christophe Loiseau, Prashant Valluri, Iain Bethune, and
  Lennon {\'O}~N{\'a}raigh.
\newblock High-performance computational fluid dynamics: a custom-code
  approach.
\newblock {\em European Journal of Physics}, 37(4):045001, 2016.

\bibitem{ding2007wetting}
Hang Ding and Peter~DM Spelt.
\newblock Wetting condition in diffuse interface simulations of contact line
  motion.
\newblock {\em Physical Review E}, 75(4):046708, 2007.

\bibitem{yue2011can}
P~Yue and JJ~Feng.
\newblock Can diffuse-interface models quantitatively describe moving contact
  lines?
\newblock {\em The European Physical Journal Special Topics}, 197(1):37--46,
  2011.

\bibitem{legendre2015comparison}
Dominique Legendre and Marco Maglio.
\newblock Comparison between numerical models for the simulation of moving
  contact lines.
\newblock {\em Computers \& Fluids}, 113:2--13, 2015.

\end{thebibliography}

\end{document}